\documentclass[%
reprint,
amsmath,amssymb,
aps,
]{revtex4-1}

\usepackage{graphicx}
\usepackage{dcolumn}
\usepackage{bm}

\begin{document}
\preprint{APS/123-QED}

\title{Discrete phase-space mappings, tomographic condition and permutation invariance }
\author{C.~Mu\~{n}oz$^1$ and A.~B.~Klimov$^{1,2}$}
\affiliation{\\$^{2}$Dept. de F\'{\i}sica, Universidad de Guadalajara, 44420 Guadalajara, Mexico\\
$^{2}$Center of Quantum Optics and Quantum Information, Center for Optics
and Photonics, Departamento de F\'{\i}sica, Universidad de Concepci\'{o}n,
Casilla-160C, Concepci\'{o}n, Chile.}
\date{\today}

\begin{abstract}
We analyze different families of discrete maps\ in the N-qubit systems in the
context of the permutation invariance. We prove that the tomographic condition
imposed on the self-dual (Wigner) map is incompatible with the requirement of
the invariance under particle permutations, which makes it impossible to
project the Wootters-like Wigner function into the space of symmetric
measurements. We also provide several \textit{explicit} forms of the self-dual
mappings: a) tomographic and b) permutation invariant \ and analyze the
symmetric projection in the latter case.
\begin{description}
\item[PACS numbers]
\end{description}
\end{abstract}


\maketitle

\section{Introduction}

Phase-space methods \cite{WF,WFb,WFc} have been widely applied in quantum
mechanics both for state visualization and analysis of quantum-classical
transitions from kinematical and dynamic perspectives. According to this
approach quantum states are mapped into distributions on some manifold, which
is associated with a "classical" phase-space. The structure of the mapping as
well as their principal characteristics depend on the symmetry of a given
quantum system. The basic requirement for any meaningful phase-space
representation is the covariance under an appropriate group of transformations
(which is one of the fundamental Stratonovich-Weyl conditions \cite{SU2},
\cite{SU2c}, \cite{WF2g}). In the case of continuous symmetries and when the
operators from the group representation act irreducibly in the Hilbert space
of the quantum system, the phase-space manifold can be constructed as a
certain quotient space and has an intimate relation to the set of coherent
states \cite{WF2} - \cite{SU2b}. Then, a systematic procedure for the
so-called $s$-parametrized phase-space mapping can be suggested at least for
some type of dynamic symmetry groups \cite{WF2g, WF2h, SUN}.

The situation is essentially more involved in case of discrete systems.
Although several approaches for representation of states of a generic
$d$-dimensional system in a discrete lattice were considered \cite{DM}, an
explicit self-consistent map possessing all the required properties can be
constructed only when $d=p^{N}$, where $p$ is a prime number \cite{DFW1}.
Then, a discrete $p^{N}\times p^{N}$ grid, playing the role of the phase-space
$\mathcal{M}_{p^{N}}$, possesses the same basic geometric properties as an
ordinary plane and allows a direct association of states with specific
geometrical structures \cite{DFW1}, \cite{GS} related to the notion of
mutually unbiased bases \cite{MUB}. A set of dual maps from the Hilbert space
$\mathcal{H}_{p^{N}}$ to $\mathcal{M}_{p^{N}}$ can be introduced in a similar
way as in the continuous case \cite{DFW2}, \cite{klimov06}, \cite{ruzzi}. It
is worth noting, that in this approach only the "boundary" maps, corresponding
to the familiar $P$ and $Q$ functions \cite{SU2c} - \cite{WF2c}, are uniquely
defined. All the other maps still can be "refined" by imposing additional
conditions to the standard Stratonovich-Weyl requirements. One possible
condition could be the marginal reduction, i.e. that summing the image of the
density matrix in $\mathcal{M}_{p^{N}}$ (a quasidistribution function) along a
set of points associated with a given state one obtains the probability
distribution in this state. This requirement gives a clear geometric
interpretation of the discrete map and is also known as a tomographic condition.

Regrettably, none of the discrete phase-space representations is efficient for
the visualization of states of large compound (many-particle) quantum systems
\cite{DCS}. It is mainly related to the ordering problem on $\mathcal{M}%
_{p^{N}}$, but is also connected to the classical indistinguishability between
irreducible degenerated subspaces (subspaces of the same dimensions that
appear in the decomposition of $\mathcal{H}_{p}^{\otimes N}$). The use of the
so-called projected $Q$-function for the analysis of N-qubit systems,
$d=2^{N}$ was recently proposed in \cite{macros}. Such a function is defined
in a three-dimensional discrete $N\times N\times N$ space and contains full
and non-redundant information about results of measurements of any invariant
under particle permutations observable in an arbitrary (not necessarily
symmetric) state. Unfortunately, being a positive image of the density matrix,
the projected $Q$-function is not convenient for graphical representation of
the interference pattern and it would be desirable to find a symmetric map
allowing to distinguish between coherent and incoherent superpositions.

In this paper we analyze the $s$-parametrized family of discrete maps for
N-qubits systems from the perspective of projection into the space of
symmetric measurements and discuss several possibilities of constructing
discrete maps according to established invariance properties.

In Sec.II we briefly recall the concept of $s$-parametrized discrete mappings.
In Sec.III we provide explicit forms for discrete mappings under different
symmetry conditions and prove the inconsistency between the tomographic and
permutation-invariance requirements. We also discuss the projected form of
self-dual mappings invariant under particle permutations.

\section{Discrete phase-space and discrete mappings}

Here we will be interested in transformations induced by elements of the
generalized Pauli group $\mathcal{P}^{N}$ acting in the N-qubit Hilbert space
$\mathcal{H}_{2^{N}}$ $=\{|\kappa\rangle$, $\kappa\in\mathbb{F}_{2^{N}}\}$. In
this case, the discrete phase space (DPS) is a $2^{N}\times2^{N}$ grid, which
points\emph{ }$(\alpha,\beta)$, $\alpha,\beta\in\mathbb{F}_{2^{N}}$ label
elements of a monomial operational basis \cite{Schwinger}, \cite{stabilizers}
$Z_{\alpha}X_{\beta}$, being
\begin{align}
Z_{\alpha}  &  =\sum_{\alpha}\chi(\alpha\kappa)|\kappa\rangle\langle
\kappa|,\quad X_{\beta}|\kappa\rangle=\sum_{\beta}|\kappa+\beta\rangle
\langle\kappa|,\label{ZX}\\
\chi(\alpha)  &  =(-1)^{tr(\alpha)},\;tr(\alpha)=\sum_{i=1}^{N-1}\alpha
^{2^{i}},
\end{align}
generators of $\mathcal{P}^{N}$, $Z_{\alpha}X_{\beta}=\chi(\alpha
\kappa)\,X_{\beta}Z_{\alpha}$. In terms of expansion coefficients of
$\alpha\in\mathbb{F}_{2^{N}\text{ }}$in a self-dual basis $\{\theta
_{1},...,\theta_{N}\}$, $tr(\theta_{i}\,\theta_{j})=\delta_{ij}\,$,
\begin{equation}
\alpha=\sum_{i=1}^{N}a_{i}\,\theta_{i}\,,\;a_{i}\in\mathbb{Z}_{2}, \label{ai}%
\end{equation}
one can associate $Z_{\alpha}$ and $X_{\beta}$ with N-particle operators%

\[
Z_{\alpha}=\sigma_{z}^{a_{1}}\otimes...\otimes\sigma_{z}^{a_{N}},\quad
X_{\beta}=\sigma_{x}^{b_{1}}\otimes...\otimes\sigma_{x}^{b_{N}},
\]
where $\sigma_{z}=|0\rangle\langle0|-|1\rangle\langle1|$, $\sigma
_{x}=|0\rangle\langle1|+|1\rangle\langle0|$, acting in $\mathcal{H}_{2^{N}}$
$=\mathcal{H}_{2}^{\otimes N}=\{|\kappa\rangle=|k_{1},...,k_{N}\rangle
,k_{i}\in\mathbb{Z}_{2}\}$. In the N-tuple $|k_{1},...,k_{N}\rangle$, where
$k_{i}$ are expansion coefficients of $\kappa$ in the self-dual basis, each
qubit is then associated with a particular element of the basis:
qubit$_{i}\Leftrightarrow\theta_{i}$. This DPS (isomorphic to a product of
two-dimensional discrete torus $T^{2}\otimes T^{2}\otimes...$) is endowed with
a finite geometry \cite{DFW1} and admits a set of discrete symplectic
transformations \cite{DFW2}, \cite{klimov06}. Thus, in complete similarity
with the continuous case, the axes of the discrete phase-space are associated
with the complementary observables $Z_{\alpha}$ and $X_{\beta}$ in the sense
that any eigenstate of either one of them is a state of maximum uncertainty
with respect to the other.

An $s$-parametrized set of quasidistribution functions satisfying the standard
Stratanovich-Weyl conditions is defined through the\ following one-to-one map
\cite{DFW2}, \cite{ruzzi},%
\begin{align}
W_{f}^{(s)}\left(  \alpha,\beta\right)   &  =\mathrm{Tr}\left[  \hat{f}%
\Delta^{\left(  s\right)  }\left(  \alpha,\beta\right)  \right]
,\label{map}\\
\hat{f}  &  =\frac{1}{2^{N}}\sum_{\alpha,\beta}W_{f}^{\left(  s\right)
}(\alpha,\beta)\Delta^{\left(  -s\right)  }(\alpha,\beta).
\end{align}
where the mapping kernel has the form%

\begin{equation}
\Delta^{\left(  s\right)  }\left(  \alpha,\beta\right)  =\frac{1}{2^{N}}%
{\displaystyle\sum\limits_{\gamma,\delta}}
\chi(\alpha\delta+\beta\gamma)\left[  \left\langle \xi\right\vert
D(\gamma,\delta)\left\vert \xi\right\rangle \right]  ^{-s}D(\gamma,\delta),
\label{Ds}%
\end{equation}
here%
\begin{align}
D(\gamma,\delta)  &  =\phi(\gamma,\delta)Z_{\gamma}X_{\delta},\;\label{D}\\
\phi(\gamma,\delta)\phi^{\ast}(\gamma,\delta)  &  =1,\;\phi(0,\delta
)=\phi(\gamma,0)=1, \label{phi}%
\end{align}
are the (unitary) displacement operators and the fiducial state $\left\vert
\xi\right\rangle $ is chosen in such a way that $\left\langle \xi\right\vert
D(\gamma,\delta)\left\vert \xi\right\rangle \neq0$.

The kernel is (\ref{Ds}) normalized
\[
\sum_{\alpha,\beta}\Delta\left(  \alpha,\beta\right)  =2^{N},
\]
covariant
\begin{equation}
D\left(  \kappa,\lambda\right)  \Delta\left(  \alpha,\beta\right)  D^{\dagger
}\left(  \kappa,\lambda\right)  =\Delta\left(  \alpha+\kappa,\beta
+\lambda\right)  , \label{cc}%
\end{equation}
and in addition it is Hermitian, $\Delta\left(  \alpha,\beta\right)
=\Delta^{\dagger}(\alpha,\beta)$, if the phase (\ref{phi}) satisfies the
condition,
\begin{equation}
\phi^{2}\left(  \gamma,\delta\right)  =\chi\left(  \gamma\delta\right)  ,
\label{phi_c22}%
\end{equation}
which also leads to the unitarity of the displacement operator, $D^{\dag
}\left(  \gamma,\delta\right)  =D\left(  \gamma,\delta\right)  $.

The overlap relation
\[
\mathrm{Tr}\left(  \Delta^{(s)}(\alpha,\beta)\Delta^{(-s)}(\alpha^{\prime
},\beta^{\prime})\right)  =2^{N}\delta_{\alpha\alpha^{\prime}}\delta
_{\beta\beta^{\prime}},
\]
is automatically fulfilled and allows the evaluatation of the trace of a
product in the form of a convolution,
\[
\mathrm{Tr}(\hat{f}\hat{g})=2^{N}\sum_{\alpha,\beta}W_{f}^{\left(  s\right)
}(\alpha,\beta)W_{g}^{\left(  -s\right)  }(\alpha,\beta).
\]

\section{Symmetries of DPS mapping}

The representation of an N-qubit state $\rho$ in DPS by any of $W_{\rho
}^{\left(  s\right)  }(\alpha,\beta)$ has an important drawback: while for a
small number of particles the plot of quasidistributions is representative, it
becomes extremely involved and is practically useless for analysis of quantum
states when $N\gg1$ \cite{DCS}. In part it is a consequence of the absence of
a natural ordering of elements of $\mathbb{F}_{2^{N}}$. In addition, the
central limit theorem is not directly applicable to the distributions labeled
by N-tuples (representations of $\mathbb{F}_{2^{N}}$) $\{(a_{1},...,a_{N})\,,$
$a_{i}\in\mathbb{Z}_{2}\}$. This explains an essentially smaller number (with
respect to the continuous case) of applications of the discrete phase-space
representations in many-body quantum mechanics.

\subsection{Permutation-invariant $s=\pm1$ mapping}

Nevertheless, this problem can be fixed if the available set of measurements
is restricted only to symmetric observables $\{\hat{S}\}$, i.e. invariant
under particle permutations, $\hat{S}=\hat{\Pi}_{ij}\hat{S}\hat{\Pi}_{ij}$,
$,i,j=1,...N$, where $\hat{\Pi}_{ij}$ is the permutation operator
\cite{macros}. It results that if the fiducial state $\left\vert
\xi\right\rangle $ in (\ref{Ds}) is permutation-invariant (i.e. it is a spin
coherent state), and $\left\langle \xi\right\vert D(\gamma,\delta)\left\vert
\xi\right\rangle \neq0$, the image $W_{S}^{\left(  \pm1\right)  }(\alpha
,\beta)$ of any symmetric operator $\hat{S}$, is a function only of the
(permutation) invariants constructed on the phase-space coordinates
$(\alpha\mathbf{,}\beta)$
\begin{align}
h\left(  \alpha\right)   &  =\sum_{i=0}^{N}a_{i},\quad h\left(  \beta\right)
=\sum_{i=0}^{N}b_{i},\label{h}\\
h(\alpha\mathbf{+}\beta)  &  =\sum_{i=0}^{N}\{a_{i}+b_{i}\},\nonumber
\end{align}
where $\{a_{i}+b_{i}\}$ means sum $\operatorname{mod}2$, and $0\leq h\left(
\kappa\right)  \leq N$. In other words, $W_{S}^{(\pm1)}(\alpha,\beta)$ are
permutation-invariant functions of the phase-space coordinates, where $\alpha$
and $\beta$ are considered as N-tuples, $\alpha=(a_{1},...,a_{N}),$ according
to (\ref{ai}).

Thus, the full information about the results of measurements of any symmetric
observable in an (arbitrary) N-qubit state $\rho$ is contained in the
projection of $W_{\rho}^{\left(  \pm1\right)  }(\alpha,\beta)$ into the 3
dimensional space spanned by $h\left(  \alpha\right)  $, $h\left(
\beta\right)  $, $h(\alpha\mathbf{+}\beta)$ (space of symmetric measurements),%
\begin{equation}
\tilde{W}_{\rho}^{(  \pm1)  }\left(  m,n,k\right)  =\sum
_{\alpha\mathbf{,}\beta}W_{\rho}^{(  \pm1)  }\left(  \alpha
\mathbf{,}\beta\right)  \delta_{m,h\left(  \alpha\right)  }\delta_{n,h\left(
\beta\right)  }\delta_{k,h\left(  \alpha\mathbf{+}\beta\right)  } \label{Wp}%
\end{equation}
While the distribution $W_{\rho}^{\left(  1\right)  }\left(  \alpha
\mathbf{,}\beta\right)  $ (corresponding to the $P$-function) becomes quite
singular for large number of qubits, $W_{\rho}^{\left(  -1\right)  }\left(
\alpha\mathbf{,}\beta\right)  $ (the $Q$-function) tends to a smooth
distribution when $N\gg1$ and is very convenient for analysis of quantum
states in the macroscopic limit \cite{macros}. It is worth noting here that
\begin{equation}
\Delta^{\left(  -1\right)  }(\alpha,\beta)=|\alpha\mathbf{,}\beta
\rangle\langle\alpha\mathbf{,}\beta|, \label{D-1}%
\end{equation}
where $|\alpha,\beta\rangle=D(\alpha,\beta)|\xi\rangle\,$\ are the so-called
discrete coherent states, that form an informational complete set of POVMs
when $\left\langle \xi\right\vert D(\gamma,\delta)\left\vert \xi\right\rangle
\neq0$.

\subsection{Covariant (Wigner) mapping}

Unfortunately, the projection $\tilde{W}_{\rho}^{\left(  -1\right)  }$ does
not distinguish very well between coherent and incoherent superpositions due
to the typical (for the $Q$-function form) of the mapping kernel (\ref{D-1}).
For instance, for the GHZ state $\left\vert GHZ\right\rangle =$ $\sim
\left\vert 0...0\right\rangle +\left\vert 1...1\right\rangle =\left\vert
0\right\rangle +\left\vert 1\right\rangle $, where in the last equation
$0,1\in\mathbb{F}_{2^{N}}$ one has
\begin{eqnarray}
&&\tilde{W}_{GHZ}^{\left(  -1\right)  }\left(  m,n,k\right)  =\frac
{R_{mnk}\left\vert \xi\right\vert ^{N}}{2\left(  1+\left\vert \xi\right\vert
^{2}\right)  ^{N}}\nonumber\\&&\times
\left[  \left\vert \xi\right\vert ^{N-2n}+\left\vert
\xi\right\vert ^{-N+2n}+2\left(  -1\right)  ^{m}\cos\left(  \frac{\pi}%
{4}\left(  N-2n\right)  \right)  \right],\nonumber\\&&
\label{GHZ Q}%
\end{eqnarray}
here%
\begin{equation}
R_{mnk}=\frac{N!}{\left(  \frac{n+m-k}{2}\right)  !\left(  \frac{n-m+k}%
{2}\right)  !\left(  \frac{m-n+k}{2}\right)  !\left(  N-\frac{n+m+k}%
{2}\right)  !}, \label{R}%
\end{equation}
and the interference described by the last term in (\ref{GHZ Q}) is negligible
compared with the principal maxima located at $(N,N,N\pm N/\sqrt{3})/2$.

In the continuous case it is known that the appropriate representation, that
"sees" the interference pattern is provided by the Wigner function, defined as
a self-dual image of the density matrix (when the same type of mapping is used
both for the density operator and for the observables in order to compute
average values by convoluting corresponding symbols). In addition, the
continuous analog of the kernel (\ref{Ds}) for $s=0$ possesses another
important property: integration of the Wigner function along a strip in
phase-space gives the marginal probability associated to the corresponding area.

In the standard construction \cite{DFW1}, the straight lines,%
\begin{equation}
\beta=\xi\alpha+\nu. \label{line}%
\end{equation}
in DPS can be associated with (appropriately ordered \cite{klimov06})
eigenstates of sets of commuting monomials $\{X_{\xi\alpha}Z_{\alpha}\}$.

The discrete self-dual map (\ref{map}), $s=0$, only guarantees that summing
the Wigner function along axes $\alpha=0$ and $\beta=0$ leads to the correct
projections on the logical basis $|\kappa\rangle$ and the dual basis
$|\tilde{\kappa}\rangle$ respectively, where $|\tilde{\kappa}\rangle$ are
eigenstates of $X_{\beta}$ operators. The requirement that summing along any
line (\ref{line}) gives the marginal probability distribution for the
observable associated with that line, is an additional condition. This
so-called \textit{tomographic condition} restricts the possible choices of the
phase (\ref{phi}) of the displacement operator (\ref{D}). It is worth noting
here, that the kernels $\Delta^{\left(  \pm1\right)  }(\alpha,\beta)$ do not
depend on this phase.

A convenient way for constructing the eigenstates $\{|\psi_{\nu}^{\xi}%
\rangle\}$ of a commuting set $\{X_{\xi\alpha}Z_{\alpha}\}$ associated to the
line (\ref{line}) is to use the rotation operator $V_{\xi}$,
\begin{equation}
V_{\xi}Z_{\alpha}V_{\xi}^{\dag}\sim Z_{\alpha}X_{\xi\alpha},\;\left[  V_{\xi
},X_{\nu}\right]  =0,\;V_{0}=I, \label{s1}%
\end{equation}
so that
\begin{equation}
|\psi_{\nu}^{\xi}\rangle=V_{\xi}X_{\nu}|0\rangle, \label{psi g}%
\end{equation}
where $|0\rangle$ is the eigenstate of $Z_{\alpha}$ with positive eignevalues
for all $\alpha\in\mathbb{F}_{2^{N}}$. The rotation operator expanded in the
dual basis $|\tilde{\kappa}\rangle$ has the form \cite{klimov06}%
\begin{equation}
V_{\xi}=\sum\limits_{\kappa}c_{\kappa,\xi}|\widetilde{\kappa}\rangle
\langle\widetilde{\kappa}|,\quad c_{0,\xi}=1. \label{V}%
\end{equation}
where the coefficients $c_{\kappa,\xi}$ satisfy the following non-linear
recurrence equation
\begin{equation}
c_{\kappa+\alpha,\xi}c_{\kappa,\xi}^{\ast}=\chi(\xi\alpha\kappa)c_{\alpha,\xi
}, \label{ck_2}%
\end{equation}
so that the unbiasedness condition $|\langle\psi_{\nu}^{\xi}|\psi_{\nu
^{\prime}}^{\xi^{\prime}}\rangle|^{2}=2^{-N}$, $\xi\neq\xi^{\prime}$ between
the states associated to the lines (\ref{line}) with different slopes is
satisfied automatically, and the rotation of $Z_{\alpha}$ gives $V_{\xi
}Z_{\alpha}V_{\xi}^{\dag}=c_{\alpha,\xi}Z_{\alpha}X_{\alpha\xi}$.
Geometrically, the action of $V_{\xi}$ can be interpreted as rotations of the
rays $\beta=\xi\alpha$: $\beta=0\overset{V_{\xi}}{\Rightarrow}\beta=\xi\alpha
$. It is worth noting that $V_{\xi}$ do not form an abelian group and satisfy
the relation $V_{\xi}^{2}=X_{\xi^{2^{N-1}}}$.

One of the possible family of solutions of (\ref{ck_2}) is
\begin{equation}
c_{\alpha,\xi}=\left(  -i\right)  ^{h\left(  \alpha^{p}\xi^{p/2}\right)
},\;p=1,2,4,8...,2^{N-1}, \label{c g}%
\end{equation}
which can be verified by direct substitution.

The solution with $p=1$,
\begin{equation}
c_{\alpha,\xi}=\left(  -i\right)  ^{h\left(  \alpha\sqrt{\xi}\right)  },
\label{h1}%
\end{equation}
where $\sqrt{\xi}$ is the square root of $\xi$ uniquely defined on
$\mathbb{F}_{2^{N}}$, possesses an extra symmetry. In this case the rotation
of the horizontal axis to the ray with the slope $\xi=1$ ($\pi/4$ rotation) is
factorized,
\[
V_{1}=\otimes_{j}\sum\limits_{k_{j}=0,1}\left(  -i\right)  ^{k_{j}}%
|\widetilde{k_{j}}\rangle\langle\widetilde{k_{j}}|,
\]
and in addition the transformation%

\[
V_{1}Z_{\alpha}V_{1}^{\dag}=\left(  -i\right)  ^{h\left(  \alpha\right)
}Z_{\alpha}X_{\alpha},
\]
produces no phase during the conversion of $Z_{\alpha}=\sigma_{z}^{\alpha_{1}%
}\otimes...\otimes\sigma_{z}^{\alpha_{N}}$ into $\left(  -i\right)  ^{h\left(
\alpha\right)  }Z_{\alpha}X_{\alpha}=\sigma_{y}^{\alpha_{1}}\otimes
...\otimes\sigma_{y}^{\alpha_{N}}$.

It is worth noting that another solution of Eq. (\ref{ck_2}) is closely
connected to the so-called graph-state formalism \cite{graph} and has the form%
\begin{equation}
c_{\alpha,\xi}=(\pm i)^{\mathbf{\alpha}^{\intercal}\Gamma\mathbf{\alpha}},
\label{c graph}%
\end{equation}
where $\mathbf{\alpha}^{\intercal}=\left[  \alpha_{1},...,\alpha_{N}\right]  $
and $\Gamma_{pq}=\left[  tr\left(  \xi\theta_{p}\theta_{q}\right)  \right]  $
is the adjacency matrix of the graph (with loops) corresponding to the ray
$\beta=\xi\alpha$.

\subsection{Tomographic condition and permutation invariance}

The imposition of the tomographic condition%

\begin{equation}
\frac{1}{2^{N}}%
{\displaystyle\sum\limits_{\alpha,\beta}}
W_{\rho}^{(0)}\left(  \alpha,\beta\right)  \delta_{\beta,\xi\alpha+\mu
}=\langle\psi_{\nu}^{\xi}|\rho|\psi_{\nu}^{\xi}\rangle, \label{TC}%
\end{equation}
leads to the following relation between the coefficients of the rotation
operator (\ref{V}) and the phase of the displacement operator (\ref{D}).
\begin{equation}
\phi\left(  \tau,\upsilon\right)  =c_{\tau,\tau^{-1}\upsilon}. \label{phi_gen}%
\end{equation}
In this case the symbol of the state $|\psi_{\nu}^{\xi}\rangle$ is just a
straight line (\ref{line}),
\begin{equation}
W_{|\psi_{\nu}^{\xi}\rangle}^{(0)}(\alpha,\beta)=\delta_{\beta,\xi\alpha+\nu},
\label{WT line}%
\end{equation}
and the kernel $\Delta^{\left(  0\right)  }(\alpha,\beta)$ acquires the form
of the sum of projectors on the lines crossing at the phase-space point
$(\alpha,\beta)$ \cite{DFW1},
\begin{equation}
\Delta^{\left(  0\right)  }(\alpha,\beta)=|\tilde{\alpha}\rangle\langle
\tilde{\alpha}|+\sum_{\xi,\nu}\delta_{\beta,\xi\alpha+\nu}|\psi_{\nu}^{\xi
}\rangle\langle\psi_{\nu}^{\xi}|-I\,. \label{wl}%
\end{equation}
In particular, the solution (\ref{c g}) leads to the following phase%

\begin{equation}
\phi\left(  \alpha,\beta\right)  =c_{\alpha,\alpha^{-1}\beta}=\left(
-i\right)  ^{h\left(  \alpha^{p/2}\beta^{p/2}\right)  },\;p=1,2,...,2^{N-1}%
. \label{phase}%
\end{equation}
In the simplest (and the most symmetric) case, $p=1$, when
\[
\phi\left(  \alpha,\beta\right)  =\left(  -i\right)  ^{h\left(  \sqrt
{\alpha\beta}\right)  },
\]
the Wigner function $W_{\rho}^{(0)}\left(  \alpha,\beta\right)  $ has the
following form for

a) GHZ-state
\begin{eqnarray}
W_{GHZ}^{(0)}\left(  \alpha,\beta\right)  &&=\frac{1}{2} \delta_{\beta,1}%
+\frac{1}{2}\delta_{\beta,0}\nonumber\\&&
+\frac{1}{2^{N}}\chi\left(  \alpha\right)
\operatorname{Re}\left[  \left(  1-i\right)  ^{N}i^{h\left(  \sqrt{\beta
}\right)  }\right]  ,
\end{eqnarray}
where the last term clearly represents the interference, practically absent,
for instance, in $W_{GHZ}^{(-1)}\left(  \alpha,\beta\right)  $,
Eq.(\ref{GHZ Q});

b) W-state (the Dicke state with one excitation), which in terms of
$\mathbb{F}_{2^{N}}$ elements can be conveniently represented as $\left\vert
W\right\rangle =N^{-1/2}\sum_{i=1}^{N}\left\vert \theta_{i}\right\rangle $,
here $\theta_{i}$ are elements of a self-dual basis,%
\begin{eqnarray}
W_{W}^{(0)}\left(  \alpha,\beta\right)  &&=\frac{1}{N}%
{\displaystyle\sum\limits_{p=1}^{N}}
\delta_{\beta,\theta_{p}}\nonumber\\&&
+\frac{\left(  1-i\right)  ^{N}}{2^{N}N}%
{\displaystyle\sum\limits_{p\neq q}^{N}}
\left(  -1\right)  ^{\alpha_{p}+\alpha_{q}}i^{h\left(  \sqrt{\frac
{\beta+\theta_{p}}{\theta_{q}+\theta_{p}}}\right)  },
\end{eqnarray}
where $\alpha_{p}$ in the interference term are components of $\alpha$ in the
self-dual basis $\{\theta_{i}\}$.

c) The Wigner function of SU(2) coherent states $\left\vert \zeta\right\rangle
$ is in general fairly complicated, except for equatorial states, when
$\zeta=1$ and $\left\vert \zeta=1\right\rangle =2^{-N/2}\sum_{\kappa
}\left\vert \kappa\right\rangle $, here $\left\vert \kappa\right\rangle $ is
the logical basis,%

\[
W_{\left\vert \zeta=1\right\rangle }\left(  \alpha,\beta\right)
=\delta_{\alpha,0}.
\]

\subsubsection{Symmetric Wigner mapping}

In order to construct a self-contained projection of $W_{\rho}^{\left(
0\right)  }(\alpha,\beta)$ on the space of symmetric measurements similar to
(\ref{Wp}), the map $\Delta^{(0)}(\alpha,\beta)$ should satisfy the basic
condition that the symbol $W_{S}^{(0)}(\alpha,\beta)$ of any symmetric
operator $\hat{S}$ is a permutation-invariant function of the phase-space
coordinates. This condition requires invariance of $\Delta^{(0)}(\alpha
,\beta)$ under particle permutations,
\begin{equation}
\hat{\Pi}_{ij}\Delta^{(0)}(\alpha,\beta)\hat{\Pi}_{ij}=\Delta^{(0)}%
(\alpha,\beta), \label{W0p}%
\end{equation}
and is fulfilled only when the phase (\ref{phi}) is an invariant function
under \textit{the same} permutations of $\alpha$ and $\beta.$

\textit{Theorem}: There does not exist a complete set of permutation-invariant
phases $\phi\left(  \alpha,\beta\right)  $ satisfying the condition
(\ref{phi_gen}).

Proof: The relation (\ref{phi_gen}) leads to the following recurrence equation
for the phase $\phi\left(  \alpha,\beta\right)  $%

\[
\phi\left(  \alpha+\beta,\alpha\xi+\beta\xi\right)  =\chi(\alpha\beta\xi
)\phi\left(  \alpha,\alpha\xi\right)  \phi\left(  \beta,\beta\xi\right)  .
\]

Let us suppose that $\phi\left(  x,y\right)  $ is invariant under a
permutation of $x$ and $y$, then $\chi(\alpha\beta\xi)=\chi((\alpha
)(\beta)(\alpha\xi)(\beta\xi))$ (here we have used the property $tr(\alpha
)=tr(\alpha^{2})$) must be also invariant under \textit{the same} permutation
of $\alpha,\beta,\beta\xi$ and $\alpha\xi$. Nevertheless, a permutation of
$r$-th and $s$-th qubits does not leave invariant $\chi(\alpha\beta\xi)$ when
$\alpha=\theta_{p}^{2}$, $\beta=\theta_{p}+\theta_{q}$, $\xi=(\theta
_{r}+\theta_{s})^{-1}$ such that $tr\left(  \theta_{r}\theta_{p}^{2}\right)
=tr\left(  \theta_{s}\theta_{p}^{2}\right)  $ for any $q$ satisfying $q\neq
p\neq r\neq s$ (here $\theta_{j}$ are elements of a self-dual basis). In fact,
taking into account that under permutation of $r$-th and $s$-th qubits the
N-tuple $\alpha$ is transformed into $\alpha^{\prime}=\alpha+\varepsilon
tr\left(  \alpha\varepsilon\right)  $, $\varepsilon=\theta_{r}+\theta_{s}$ we
observe that%

\[
\chi\left(  \left[  \alpha^{\prime}\right]  \left[  \beta^{\prime}\right]
\left[  \alpha\varepsilon^{-1}\right]  ^{\prime}\left[  \beta\varepsilon
^{-1}\right]  ^{\prime}\right)  =-\chi\left(  \left[  \alpha\right]  \left[
\beta\right]  \left[  \beta\varepsilon^{-1}\right]  \left[  \alpha
\varepsilon^{-1}\right]  \right)
\]
which means that for these values $\alpha,\beta$ and $\xi$ the phase
$\chi\left(  \left[  \alpha\right]  \left[  \beta\right]  \left[  \beta
\xi\right]  \left[  \alpha\xi\right]  \right)  $ is not invariant under the
permutation $\varepsilon$.

Thus, one can not project the Wigner map (\ref{wl}) into the space of
symmetric measurements in a manner similar to (\ref{Wp}).

\subsubsection{Permutation-invariant phase}

As we have proved, the tomographic condition is incompatible with the
permutation invariance of the phase (\ref{phi}). Nevertheless, withdrawing
this requirement and demanding only the Hermiticity of the map (\ref{cc}) one
can find multiple permutation-invariant solutions of Eq.(\ref{phi_c22}). The
simplest one is an arbitrary distributed (on $\pm$ signs) set of square
roots,
\begin{equation}
\phi\left(  \alpha,\beta\right)  =\pm\sqrt{\chi\left(  \alpha\beta\right)  }.
\label{phi s1}%
\end{equation}
Also, the phase $\phi\left(  \alpha,\beta\right)  $ can be represented
directly as a function of the lengths (\ref{h}), $\phi\left(  \alpha
,\beta\right)  =\phi(h\left(  \alpha\right)  ,h\left(  \beta\right)  ,h\left(
\alpha+\beta\right)  )$. The simplest form of such phase is%

\begin{equation}
\phi\left(  \alpha,\beta\right)  =(-1)^{f(\alpha,\beta)}i^{\frac{h\left(
\alpha\right)  +h\left(  \beta\right)  -h\left(  \alpha+\beta\right)  }{2}},
\label{phi s}%
\end{equation}
(observe that $h\left(\alpha\right)  +h\left(  \beta\right)  -h\left(  \alpha+\beta\right)=2 \sum_{i}\alpha_{i},\beta_{i}$)
 where $f(\alpha,\beta)$ is an arbitrary permutation invariant function. If the
function $f(\alpha,\beta)$ is in addition factorizable,
\[
f(\alpha,\beta)=\sum_{i}f_{i}(\alpha_{i},\beta_{i}),
\]
the kernel $\Delta^{\left(  0\right)  }\left(  \alpha,\beta\right)  $ has a
product form,%
\begin{align}
\Delta^{\left(  0\right)  }\left(  \alpha,\beta\right)   &  =\otimes_{i}%
\Delta_{i}^{\left(  0\right)  }\left(  \alpha_{i},\beta_{i}\right)
,\label{D0}\\
\Delta_{i}^{\left(  0\right)  }\left(  \alpha_{i},\beta_{i}\right)   &
=\frac{1}{2}%
{\displaystyle\sum\limits_{\gamma_{i},\delta_{i}=0}^{1}}
(-1)^{\alpha_{i}\delta_{i}+\beta_{i}\gamma_{i}+f(\gamma_{i},\delta_{i})}i^{\gamma_{i}\delta_{i}%
}\sigma_{z}^{\gamma_{i}}\sigma_{x}^{\delta_{i}%
}.\nonumber
\end{align}

Although, the property (\ref{WT line}) is not true for the permutation
invariant map (\ref{D0}) in general, it still holds for the factorized bases
(eigenstates of the sets $\{Z_{\alpha}\}$, $\{X_{\alpha}\}$, $\{Z_{\alpha
}X_{\alpha}\}$),%

\begin{align*}
W_{\left\vert \kappa\right\rangle \left\langle \kappa\right\vert }\left(
\alpha,\beta\right)   &  =\delta_{\beta,\kappa},\\
W_{\left\vert \widetilde{\kappa}\right\rangle \left\langle \widetilde{\kappa
}\right\vert }\left(  \alpha,\beta\right)   &  =\delta_{\alpha,\kappa},\\
W_{\left\vert \psi_{\kappa}^{1}\right\rangle \left\langle \psi_{\kappa}%
^{1}\right\vert }\left(  \alpha,\beta\right)   &  =\delta_{\alpha+\beta
,\kappa},
\end{align*}
where $\left\vert \psi_{\kappa}^{1}\right\rangle =V_{1}\left\vert
\kappa\right\rangle $.

Correspondingly, the symbols of symmetric operators are permutation invariant
function, in particular, for the image of the SU(2) group element%

\[
g=\exp\left(  i\varphi S_{z}\right)  \exp\left(  i\theta S_{x}\right)
\exp\left(  i\psi S_{z}\right)  ,
\]
where $S_{x,y,z}=\sum_{i=0}^{N}\sigma_{x,y,z}^{(i)}$, being $\sigma
_{x,y,z}^{(i)}$ Pauli matrices,
\begin{gather*}
W_{g}\left(  \alpha,\beta\right)  = \cos^{N}\theta\\
\left[  e{^{i\phi+i\psi}}+i\sqrt{2}\tan\theta\cos\left(  \phi
-\psi-\pi/4\right)  \right]  ^{N-\frac{h\left(  \alpha\right)  +h\left(
\beta\right)  +h\left(  \alpha+\beta\right)  }{2}}\\
\left[  e{^{-i\phi-i\psi}}+i\sqrt{2}\tan\theta\cos\left(  \phi
-\psi+\pi/4\right)  \right]  ^{\frac{-h\left(  \alpha\right)  +h\left(
\beta\right)  +h\left(  \alpha+\beta\right)  }{2}}\\
\left[  e{^{i\phi+i\psi}}-i\sqrt{2}\tan\theta\cos\left(  \phi
-\psi-\pi/4\right)  \right]  ^{\frac{h\left(  \alpha\right)  -h\left(
\beta\right)  +h\left(  \alpha+\beta\right)  }{2}}\\
\left[  e{^{-i\phi-i\psi}}-i\sqrt{2}\tan\theta\cos\left(  \phi
-\psi+\pi/4\right)  \right]  ^{\frac{h\left(  \alpha\right)  +h\left(
\beta\right)  -h\left(  \alpha+\beta\right)  }{2}}%
\end{gather*}
held for $f(\alpha,\beta)=0$.

The Wigner function defined with permutation-invariant phases can be
faithfully mapped into the measurement space according to,
\begin{equation}
\tilde{W}_{\rho}^{\left(  0\right)  }\left(  m,n,k\right)  =\sum
_{\alpha\mathbf{,}\beta}W_{\rho}^{\left(  0\right)  }\left(  \alpha
\mathbf{,}\beta\right)  \delta_{m,h\left(  \alpha\right)  }\delta_{n,h\left(
\beta\right)  }\delta_{k,h\left(  \alpha\mathbf{+}\beta\right)  },
\label{W0pr}%
\end{equation}
so that average values of any symmetric operator $\hat{S}$ is computed as a
convolution%
\begin{equation*}
\langle\hat{S}\rangle=2^{N}\sum\limits_{m,n=0}^{N}\sum\limits_{k=\left\vert
m-n\right\vert }^{\min\left(  m+n,N,2N-m-n\right)  }\tilde{W}_{\rho}^{\left(
0\right)  }(m,n,k) \tilde{W}_{S}^{\left(  0\right)  } (m,n,k)  ,
\end{equation*}
where $\tilde{W}_{S}^{\left(  0\right)  }\left(  m,n,k\right)  $ is the Wigner
symbol of $\hat{S}$.

For instance, the Wigner function under the choice (\ref{phi s}) with
$f(\alpha,\beta)=0$, for the GHZ state has the form%

\begin{eqnarray}
W_{\left\vert GHZ\right\rangle }^{\left(  0\right)  }\left(  \alpha
,\beta\right)  &&=\frac{1}{2}\delta_{\beta,0}+\frac{1}{2}\delta_{\beta,1}\nonumber\\&&%
+\chi\left(  \alpha\right)  \operatorname{Re}\left[  \left(  1+i\right)
^{N}\left(  -i\right)  ^{h\left(  \beta\right)  }\right]\nonumber
\end{eqnarray}
which leads to the projection (\ref{W0pr})%

\begin{eqnarray}
&&\tilde{W}_{\left\vert GHZ\right\rangle }^{\left(  0\right)  }\left(
m,n,k\right)  =\frac{1}{2}\delta_{n,0}\delta_{m,k}C_{N}^{k}+\frac{1}{2}%
\delta_{n,N}\delta_{m,N-k}C_{N}^{m}\nonumber\\&&
+R_{mnk}\left(  -1\right)  ^{m+n}%
\operatorname{Re}\left[  \left(  1+i\right)  ^{N}i^{n}\right]\nonumber
\end{eqnarray}
where $R_{mnk}$ is defined in (\ref{R}) and $C_{N}^{m}$ are the Binomial
coefficients. One can observe a large interference term centered at $\left(
N/2,N/2,N/2\right)  $ while the maxima corresponding to $\left\vert
0\right\rangle $ and $\left\vert 1\right\rangle $ are located now at $\left(
N/2,0,N/2\right)  $ and $\left(  N/2,N,N/2\right)  $ respectively (compare
with (\ref{GHZ Q})). This picture is very similar to the representation of the
interference of Schrodinger cat-like states in the flat phase-space.

\section{Conclusions}

The discrete Wigner map corresponding to (\ref{map})-(\ref{Ds}) at $s=0$ can
be either permutation symmetric or satisfy the tomographic condition
(\ref{TC}). In both cases there exist multiple constructions of such mappings.
Symmetric maps allow the projection of the Wigner function into the space of
symmetric measurements and detect the interference patterns, separated from
the contribution of the incoherent terms. Nevertheless, there is no symmetric
projection of the Wootters-like Wigner function defined by the map (\ref{wl}),
which essentially limits the application of this map for analysis of large
N-qubit systems.

During the preparation of this paper we found an article by Huangjun Zhu,
Phys. Rev. Lett. \textbf{116}, 040501 (2016) where the relation of the
discrete Wigner function with permutation symmetry was analyzed in a different context.

This work is partially supported by the Grant 254127 of CONACyT (Mexico).

\end{document}